\documentstyle[aps,twocolumn]{revtex}

\begin{document}

\newcommand{\beq}{\begin{equation}}
\newcommand{\eeq}{\end{equation}}
\newcommand{\beqa}{\begin{eqnarray}}
\newcommand{\eeqa}{\end{eqnarray}}

\def\ket#1{|\,#1\,\rangle}
\def\bra#1{\langle\, #1\,|}
\def\braket#1#2{\langle\, #1\,|\,#2\,\rangle}
\def\proj#1#2{\ket{#1}\bra{#2}}
\def\expect#1{\langle\, #1\, \rangle}
\def\kpsi{\ket{\psi}}
\def\kphi{\ket{\phi}}
\def\bpsi{\bra{\psi}}
\def\bphi{\bra{\phi}}

\def\half{\frac{1}{2}}

\title{Pulsed energy-time entangled twin-photon source for 
quantum communication}

\author{J. Brendel, N. Gisin, W. Tittel and H. Zbinden}

\address{University of Geneva, Group of Applied Physics, 20, 
Rue de l'Ecole de M\'edecine, CH-1211 Geneva 4, 
Switzerland}
\date{\today} 
\maketitle
\abstract{
A pulsed source of energy-time entangled photon pairs pumped 
by a standard laser diode is
proposed and demonstrated. 
The basic states can be distinguished by their time of arrival. 
This greatly simplifies the realization of 2-photon quantum 
cryptography, Bell state analyzers, 
quantum teleportation, dense coding, entanglement swapping, 
GHZ-states sources, etc.
Moreover the entanglement is well protected during photon 
propagation 
in telecom optical fibers, opening the door to
few-photon applications of quantum communication over long distances. 
}

\vspace{1 cm}
\normalsize
%\section{Introduction}
%=====================
Quantum communication offers fascinating possibilities to the 
physicists. Some correspond to
potential applications, like quantum cryptography, others explore 
the quantum world of
entanglement, like dense coding,
entanglement swapping (entangling particules that never interact)
or teleportation (transferring the unknown quantum state from one 
particle to a distant one)
\cite{Ekert91,DensecodTh,EntSwapTh,QteleportationTh,PhysWorld}.
In recent years, quantum communication, like all the field of 
quantum information processing, 
underwent an impressive flow of theoretical ideas. The experiments, 
however, were generally far behind. 
This unbalanced situation still remains, except for the 1-qubit quantum 
cryptography case (actually, 
pseudo-1-qubit, since weak coherent light pulses mimic the qubit) 
\cite{PhysWorld}. 
There is thus a clear need
for original implementations of the general ideas. In this letter, 
we propose a compact,
robust (and low cost) source producing
energy-time entangled pairs of photons (twin-photons) 
at determined times. The source can be tuned to produce any desired 
2-qubit state, in
particular the four Bell states. Contrary to other Bell state sources 
\cite{source}, 
the basic states of our twin-photons are neither based on polarization, 
nor
on momentum, but on time bins. This allows one to separate the basic 
state easily,
without any optical element, and prevents
crosstalk during the photon propagation.

We first introduce the basic states of our qubit space. Next,
we present our source and an experimental demonstration is discussed. 
Finally, the potential of our source is illustrated by several examples.

%\section{The Poincar\'e sphere of the short-long states}
%=======================================================
To understand our source, it is useful to start with the simple device 
of Figure 1 which
can be entirely understood in terms of classical linear optics. 
First, we analyse it as a preparation device. 
Let a 1-photon pulse enter the device from the left.
Assuming the pulse duration is short compared to the arm length difference 
long-short of the
Mach-Zehnder interferometer, the output consists of two well separated 
pulses. Let us
denote them $\ket{short}$ and $\ket{long}$. They form the bases of our 
qubit space,
similar to the usual vertical $\ket{V}$ and horizontal $\ket{H}$ linear 
polarization states.
Hence the state at the output of our preparation device reads 
\beq
\alpha\ket{short}+\beta\ket{long}
\label{psi}
\eeq
The relative norm and phase of the coefficients $\alpha$ and $\beta$ are 
determined by the
coupling ratio of the beam splitter and the phase shifter, respectively. 
Hence, any
state of the 2-dimensional Hilbert space spanned by the basic states 
$\ket{short}$ and
$\ket{long}$ can be prepared. The
switch of the device recombines the pulses traveling through the short 
and the long arms
without introducing any loss. It could be replaced by a passive 50-50\% 
beam splitter, at the cost of
50\% losses. Next, the same device can be used as an analyser. Simply, 
let the 2 pulses enter
the device from the right. The switch is synchronized such that the 
pulse corresponding to the
ket $\ket{short}$ takes the long path in the interferometer and vice-versa 
for the other pulse.
Hence, at the output (left) of the analyzer, both pulses interfere. 
Depending on the 
phase shift and coupling ratio the interference is constructive or 
destructive and 
complete or incomplete, respectively, in full analogy with a 
polarization analyzer. 

The correspondence between the polarization states and the states 
obtained by superposition
of the $\ket{short}$ and $\ket{long}$ ones can be extended. For 
example, a polarization
beam splitter that separates the basic vertical and horizontal 
polarization states 
corresponds to an optical switch between the short and the long pulses.

%\section{The source}
%==================================================================
Our twin-photon source consists of a pulsed pump laser, a device
similar to that described above and a non-linear
crystal where the twin-photons are created by spontaneous parametric 
downconversion.
Each photon from the pump laser is split in
two parts by the preparation device, and the two parts pass through 
the nonlinear crystal with a
time delay. Thus, if a pump photon is split into a twin-photon, the 
time of creation of
the latter is undefined. More precisely, the preparation device 
transforms the state of 
the pump photon in a superposition 
$\alpha\ket{short}_{pump}+\beta\ket{long}_{pump}$
and the
downconversion process in the crystal transforms this state into 
\cite{pumpphoton}:
\beq
\alpha\ket{short}_{s}\otimes\ket{short}_{i}+
\beta\ket{long}_{s}\otimes\ket{long}_{i}
\label{outstate}
\eeq
This is similar to the entangled state used for Franson-type tests 
of Bell inequalities 
\cite{Franson89}. However, contrary to other sources of energy-time 
entangled photons 
\cite{Franson89,BellEnergyTime,Tittel97,Tittel98b,Tittel98a}, 
the coherence of the pump laser of our source is of no importance,
as the necessary coherence is build by the preparation interferometer.
In other words, the uncertainty of the pump photon's arrival time at 
the crystal (within the 
coherence length of the pump laser)
is replaced by the two sharp values corresponding to  
$\ket{short}$ and $\ket{long}$ which form the basis of our qubit space.
Hence, any standard laser diode, for instance, can be used as pump.
Note that the pulse duration must be shorter than the arm length 
difference of 
the interferometer.

%\section{Experimental demonstration of the twin-photon source}
%===========================================================================
Figure 2 shows the twin-photon source that we used as demontrator.
It is pumped by a standard red laser diode
(Sanyo DL-LS52, $\lambda$ = 655 nm) operated in pulsed mode 
(300 ps pulses, peak power 30 mW, repetition rate 100 MHz).
A dispersing prism P deflects any infrared
emission of the laser from the entrance of the following bulk 
optics Michelson 
interferometer. At the exit of the interferometer the laserpulses 
are
split into two pulses temporarly separated by 1.2 ns, as described 
above   
for our simple device. The aperture A guarantees that both pulses belong 
to the same spatial mode. These two pulses pass through a nonlinear 
LiNbO$_3$ crystal 
which is cut to produce wavelength degenerated twin-photons 
with a center wavelength of 1310 nm. Finally, these twin photons are 
coupled into a single mode fiber by the coupler L2, the red laser light
being blocked by the filter F. 
The inset in Fig. 2 shows the results of a first experiment
in an optical setup similar to the one proposed by Franson \cite{Franson89}.
In principle each photon is analyzed by a different
analyzer with equal path-length differences. In practice, it is simpler 
to direct
both twin-photons to the same analyzer \cite{Brendel91}.
This does not
affect the possibility to characterize our source by measuring the 
2-photon interferences
produced by the undistinguishable paths: pump photon in the short 
(long) arm and the two twin photons
in the long (short) arms. 
The analyzer is an all fiber Michelson interferometer with Faraday 
mirrors (FM) to compensate polarization fluctuations 
\cite{Tittel97,Tittel98b}. 
The path difference 
corresponds exactly (within the coherence time of the pump laser) to 
the delay produced by the first interferometer. It can be varied changing
the temperature of the whole interferometer.
The optical circulator C at the input directs the backreflected 
photons to one
detector, a second is located
at the output of the interferometer. Both detectors are passively 
quenched Germanium APD's cooled to 77 K.
We record threefold coincidences between the the two detectors and the laser
pulser within a 500 ps window.
The measured interferogram is
shown in the inset of Figure 2. The measured visibility of 84\% 
clearly demonstrates that our
source produces the quantum state (\ref{outstate}). The difference 
to the ideal 100\%
visibility is attributed mainly to the mismatch between the two 
interfering modes at the output 
of the bulk interferometer. We estimate that 
with an all fiber interferometer, a stronger laser and by gating the
detectors that a visibility of more than 95\% 
should be achievable with no more
than a few seconds integration time per data point. Note that the 
preparation and analyzer 
devices act on photons of different wavelengths.

%\section{Applications}
%====================================================================
Our twin-photon source uses standard components, is compact (in future it 
could be fully integrated on an optical chip) and is
well adapted for quantum communication over optical fiber networks. 
Indeed, the separation
between the long and short paths can be made large enough to eliminate 
all drawbacks due
to dispersion of the pulses during transmission. Moreover, polarization 
fluctuations and depolarization,
inevitable in optical fibers, have no effect on our system, as already 
demonstrated by our
long distance quantum correlation experiments \cite{Tittel98b,Tittel98a}. 
Another significant advantage of our pulsed
source is that the detectors can be opened only during the short time 
windows when photons are
expected. This allows to gate the detectors and increase the detector 
efficiency from a few
percents to tens of percents. It also opens the door for InGaAs APD 
which can work
at temperatures achievable with thermo-electric cooling, but only in 
such a gated mode
\cite{Ribordy98}.
 
%\subsection{quantum cryptography}
%==================================================================
Quantum cryptography could well be the first application of quantum 
communication. So far all
demonstrations outside the lab used the 1-photon scheme 
\cite{PhysWorld,BB84,B92,weakpulses}. 
Our source should
allow a field demonstration of quantum cryptography using the 2-photon scheme
\cite{Ekert91}. This has the advantage, in addition to elegance, 
of increasing the distance, 
since one would start with 1-photon states instead of 0.1 photon 
weak pulses.
Moreover, as illustrated by Figure 3, our
source provides a simple passive detection scheme. For each 
twin-photon, each detector
can register a photon at 3 different times 
(relative to the emission time): 
short, medium, long. Short and long counts
on Alice and Bob sides correspond to the $\{\ket{short},
\ket{long}\}$ basis and are 
100\% correlated. Medium counts correspond to the
complementary basis $\{\ket{short}\pm\ket{long}\}$ and are 
also perfectly correlated
(assuming $\varphi+\delta_A+\delta_B=0$). Note that
in the first basis, the correlation is in the detection times, 
whereas in the second basis the
correlation is between the detectors that count the photons. 
We like to emphasize the
relative simplicity of this implementation: usually, in order 
to avoid an external random number
generator and a phase modulator, two analyzers are needed on 
each side.
Here one of these is realized by simply measuring the time of 
arrival of the photons, hence 
this analyzer does not require any optical element!

%\subsection{Q teleportation, entanglement swapping and dense coding}
%====================================================================
Other fascinating possibilities of quantum communication are 
teleportation, entanglement
swapping and dense coding, as already demonstrated in laboratory 
\cite{DensecodExp,QteleportationExp,Boschi98,EntSwapExp}. Our 
source provides
means to achieve these tasks over much longer distances. 
For example, consider the setup
of Figure 4. Two independent - but synchronized - twin-photon 
sources emit photon pairs
\cite{practice}.
One element of each pair is jointy analyzed by a so-called 
Bell-state analyzer. Ideally, the
eigenstates of this analyzers are the 4 Bell states:
\beqa
\psi^\pm&=&\ket{short,long}\pm\ket{long,short} \\
\phi^\pm&=&\ket{short,short}\pm\ket{long,long} 
\eeqa
In practice however, the best analyzer that can be done using only 
linear optics separates
unambiguously between 2 of 4 Bell states, leaving the 2 others 
undistinguished \cite{Weinfurter94}.
In our case, such an optimal linear analyzer is straightforward to 
implement: a 50\%
beam splitter and 2 detectors suffice! One couldn't dream of anything 
simpler! 
Indeed, consider first the case of an input state in the space spanned by
the $\phi^\pm$ states, then both photons are detected in coincidence, 
either by one or
by both detectors (the time of
detection of these 2 photons allows to distinguish between 
the $\ket{short,short}$ and the
$\ket{long,long}$ states). Next, if the input state is $\psi^-$, 
then necessarily 
both detectors get one photon, but with a time delay. Finally, 
if the input state is
$\psi^+$, then necessarily both photons propagate to the same detector, 
again with a time delay.
The two $\psi^\pm$ states can thus be unambiguously distinguished. 
This happens with a 50\% 
probability and in these cases the photons of Bob and Charly get entangled, 
despite that they
never interact directly. This is called entanglement swapping 
\cite{EntSwapTh,EntSwapExp}. The
same basic configuration could also be used for quantum teleportation
\cite{QteleportationTh,QteleportationExp}.

%\subsection{GHZ-like states}
%===========================================================================
A (pulsed) 3-photon source emitting GHZ-states \cite{GHZ} 
(i.e. maximally entangled
triplets) could work as follows, generalizing the proposal 
\cite{3outof2}. Remove one of the detectors of figure 4
and consider the cases where the remaining detector registers a 
photon at the time $short$.
Then, provided there is one photon in each of the 3 output ports, 
these 3 photons are in the 
GHZ-state: $\ket{short,long_1,long_1}+\ket{long_2,long_2,short}$, 
where $long_j$ refers to the
long arm of the $j$th source. If $long_1$ and $long_2$ are 
sufficiently different, then
this is a GHZ-state.

Yet another analysis suggested by our source is illustrated on Figure 3. 
The symmetry
of this figure is strikingly similar to that of a GHZ-state. Of course 
in a real GHZ-state
all particle are produced at the center and propagate towards their 
analyzers, Alice, Bob and
Charly. While in our case Charly's photon is the pump photon, and 
Alice and Bob receive
the signal and idler photons. Nevertheless, the analogy can be 
pushed further:
all correlations that hold for GHZ-states, hold also for the 3 
photons of figure 3. An
advantage of such states, for applications like three party quantum 
cryptography, is that
no triple coincidence photon counting is needed, as the "third photon" 
is in the bright
laser pulse.
 
Note that our source can produce non-maximally entangled states, 
using a coupler with a
coupling ratio different from $\half$ (see figure 1). This, together 
with the possibility 
to increase the detector's efficiency by time-gating, may lead to a 
test of Bell inequality
closing the annoying detection loophole \cite{localityloophole}. 
Indeed, Eberhard has shown that, surprisingly,
non-maximally entangled states are favorable for such a test 
\cite{Eberhard93}.

%\section{Concluding remarks}\label{sec:conre}
%============================================

Several generalizations of our proposal are worth to mention. 
One could use interferometers
with more than 2 branches, providing thus entanglement of 
3-dimensional quantum objects
(qutrits) or even of higher dimensions \cite{Zukowski97}. 
Another generalization would 
be to prepare the train of
coherent pulses that pump the nonlinear crystal from a coherent 
laser beam and an electro-optical
switch. Appropriate electronic driving of such an intensity and 
phase modulator could allow
one to explore entanglement of two objects of dimensions of up 
to several hundreds.

In conclusion, a compact and simple twin-photon source has been
proposed and demonstrated and several
applications in the field of quantum communication described. 
The use of time as a basis
to encode qubits makes it possible to tailor the coherence of 
the pump beam, starting
from any convenient light source. It also allows to discriminate 
between the
basic states simply by the detection time, thus simplifying Bell 
state analyzers. 
Depolarization during
the photon propagation has no effect and polarization fluctuations 
in the analyzing
interferometers can be entirely compensated thanks to Faraday 
mirrors. All this greatly
simplifies the practical implementation of quantum cryptography, 
teleportation and other
protocols over large distances using telecom optical fibers.

%\section*{Acknowledgement}
%=========================
Stimulating discussions with Bruno Huttner and 
financial support by the Swiss Priority Programme in Optics and 
by the European TMR network
"The physics of quantum information" are acknowledged.

\newpage

\section*{Figure Captions}
%=========================
\begin{enumerate}
\item Schematic of the preparation and analyzer device 
(using optical fibers and fiber couplers). 
By adjusting the coupling
ratio $\eta$ of the coupler (beam splitter) and the phase 
$\varphi$ of the phase shifter, 
any superposition
(\ref{psi}) of the basic states $\ket{short}$ and $\ket{long}$ 
can be prepared and analyzed
($\frac{|\alpha|^2}{|\beta|^2}=\frac{1-\eta}{\eta}$). The arm
length difference $\delta t$ 
of this Mach-Zender interferometer should be much longer than the pulse
duration. The (optional) 
optical switch allows to couple or separate the basic states 
without losses.

\item Schematic of the experiment to demonstrate the twin-photon source. 
The measured 
2-photon interference  visibility of 84\% establishes the non-classical 
nature of the
2-photon field (see inset).

\item Implementation of the twin-photon source for quantum cryptography. 
Alice and Bob
require only one interferometer each and no modulators (see text). 
Note also the analogy between
the 3-particle GHZ-states and the pump+twin-photon state of this 
configuration.

\item Application of the two twin-photon source for entanglement 
swapping or for quantum
teleportation. If Alice's detectors find the 2 photons out of 
coincidence, then Bob and
Charly's photon get entangled (entanglement swapping). If 
furthermore Charly measures
his photon (with the analyzer of figure 1), 
then he effectively prepares the twin-photon
that is in Alice hands and Alice measurement transfers 
(teleports) that state to Bob's
photon (up the a unitary transformation which is determined by 
Alice results). The Bell
analyzer (for 2 out of the 4 Bell states) is realized by a simple coupler.

\end{enumerate}


\begin{references}
\bibitem{pumpphoton} Actually, the pump pulse consists of 
        many photons in a coherent
       (i.e. classical) state. However, with high probability, 
       at most one of them is
       downconverted into twin-photons. Hence, we effectively 
       post-select one of the
       pump photons.
\bibitem{practice} In practice the same laser could pump both sources. 
        The main experimental
       difficulty would presumably be the requirement that the pulse 
       duration is shorter than the
       coherence time of the downconverted photon.
\end{references}
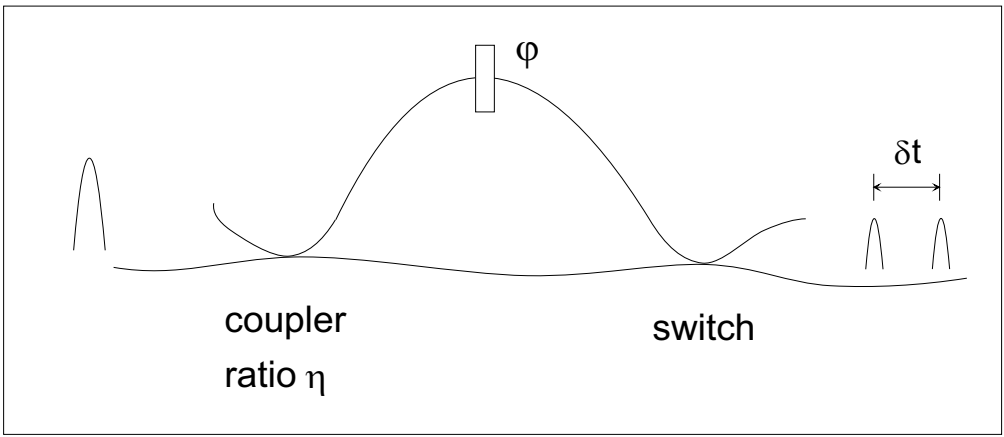
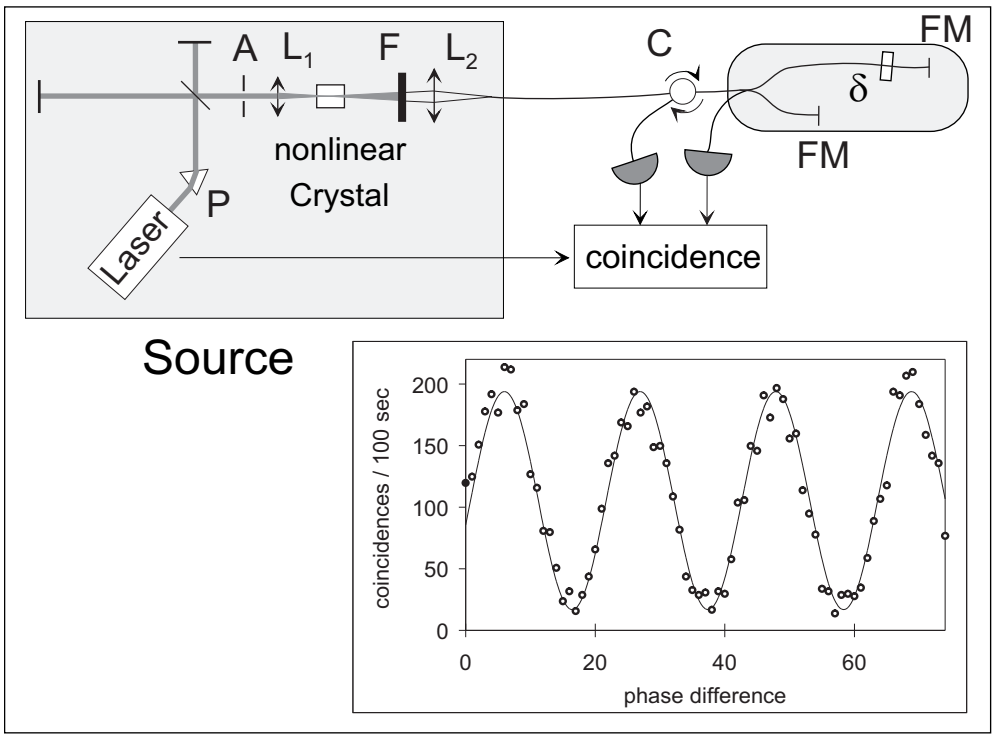
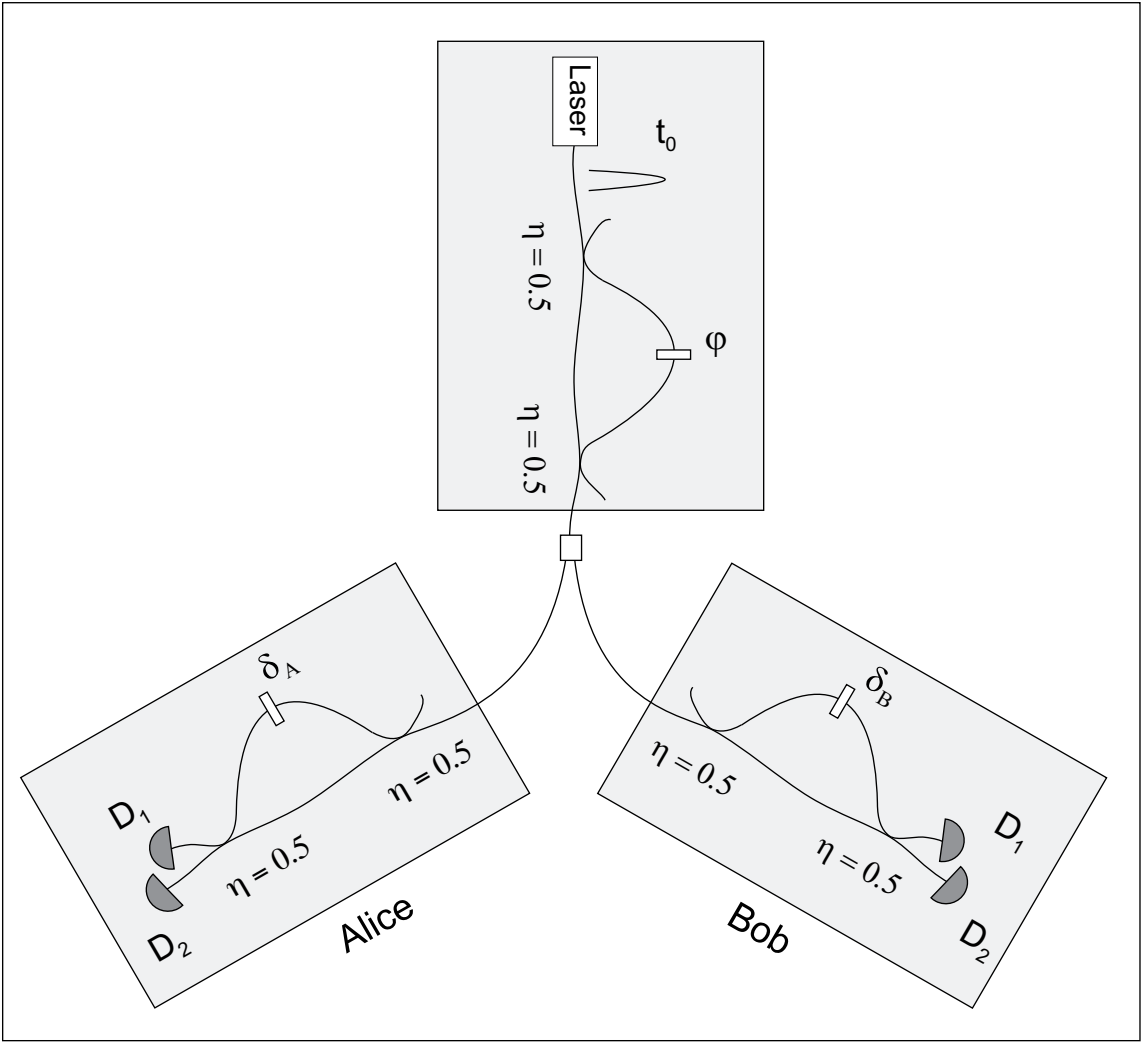
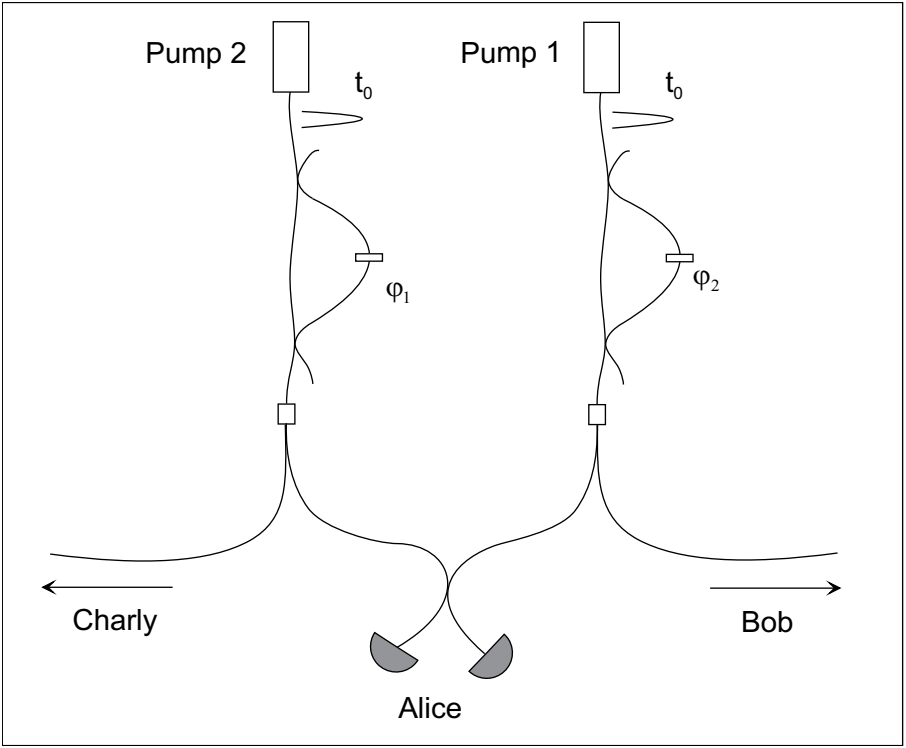
\end{document}